# Triple GEM operation in compressed He and Kr

A. Bondar, A. Buzulutskov *, L. Shekhtman, V. Snopkov, A. Vasiljev

*Budker Institute of Nuclear Physics, 630090 Novosibirsk, Russia*

**Abstract**

We study the performance of the triple GEM (Gas Electron Multiplier) detector in pure noble gases He and Kr at high pressures, varying from 1 to 15 atm. The operation in these gases is compared to that recently studied in Ne, Ar and Xe. It turned out that light noble gases, He and Ne, have superior performances: the highest gain, approaching $10^5$, and an unusual gain dependence on pressure. In particular, the maximum gain in He and Ne does not decrease with pressure, in contrast to Ar, Kr and Xe. These results are relevant for understanding basic mechanisms of electron avalanching in noble gases and for applications in cryogenic particle detectors, X-ray imaging and neutron detectors.

*PACs:* 29.40C; 07.85.F; 34.80.M

## 1. Introduction

There is a growing interest in developing gas avalanche detectors capable of operating in pure noble gases in a wide pressure range. On one hand, this interest is induced by the development of sealed gas photomultipliers [1,2], filled with pure noble gases. Such a filling should prevent the photocathode degradation under avalanche conditions. On the other hand, this interest is motivated by recent suggestions to employ light noble liquids He and Ne and heavy noble liquid Xe as detection media for solar neutrinos [3] and dark matter [4], correspondingly. A detector of choice might be a double-phase device [5], where the ionization produced in the liquid by a particle is detected in the gas phase with the help of a Gas Electron Multiplier (GEM) [6] operated at cryogenic temperatures. GEM-based detectors are particularly promising for these applications due to the recently discovered capability of multi-GEM structures to operate in pure noble gases at high gains [7].

It should be remarked that the operation at cryogenic temperatures and atmospheric pressure is equivalent to that at room temperatures and high pressures. Indeed, the gas density is a reciprocal function of the temperature at a given pressure and just the gas density is the parameter defining electron avalanche characteristics. Little is known, however, about GEM performance at high pressures [8,9] and in particular in compressed pure noble gases. So far, all investigators observed that the maximum GEM gain rapidly decreased with pressure.

---

* Corresponding author. Fax: 7-3832-342163. Phone: 7-3832-394833.
Email: buzulu@inp.nsk.su

Recently, we have studied the operation of the triple GEM detector in Ne, Ar and Xe [10] in the pressure range of 1-10 atm. The interesting observation was that Ne showed quite different pressure dependence of gain characteristics as compared to Ar and Xe, indicating that the gas amplification mechanism in light noble gases may be different at high pressures.

In this work, we further study the performance of the triple GEM detector in other two noble gases, He and Kr, at pressures reaching 15 atm. We confirm that the operation in light noble gases is substantially different from that of heavy noble gases: the performance in He and Ne turned out to be superior in terms of gain-voltage characteristics and their dependence on pressure. A possible explanation for such behaviour is presented.

## 2. Experimental setup and procedure

The experimental setup and procedure were similar to that used in [10] (see Fig.1). Three GEM foils (50 µm thick Kapton, 80 µm hole diameter at a 140 µm pitch, 28×28 mm$^2$ active area) and a printed-circuit-board (PCB) anode were mounted in cascade inside a stainless-steel vessel. The GEMs were produced at CERN workshop. The drift gap (between the cathode and the 1$^{st}$ GEM), transfer gaps (between the GEMs) and induction gap (between the last GEM and the anode) were 3, 1.6 and 1.6 mm, respectively. The detector was filled with He, Ne, Ar, Kr or Xe of 99.99% purity; it could safely operate at pressures reaching 15 atm.

The detector was irradiated with an X-ray tube, having a molybdenum target, through a 1 mm thick Al window. The voltage and current supplied to the tube were in the range of 20-30 kV and below 60 µA, respectively. At these voltages the tube radiation spectrum has two peaks: at the molybdenum characteristic lines $K_\alpha$ at 17.4 keV and $K_\beta$ at 19.7 keV. In Ne, Ar, Kr and Xe the primary ionization in the drift gap is produced by photoelectric absorption of X-ray photons by the gas molecules, while in He it is mostly produced by an X-ray-induced electron emission from the GEM electrode. The latter process will be discussed in more detail in sect. 4.

GEM electrodes were biased through a resistive high-voltage divider, as shown in Fig.1. Two voltage divider configurations, symmetrical and asymmetrical, were used. The dividers were optimized in such a way as to obtain the maximum gain in an appropriate gas. In the asymmetrical divider, the voltage across a single GEM increased towards the last GEM. In the symmetrical divider, the voltage applied to each GEM was uniform; in addition, the voltage across the induction gap was doubled compared to that of the asymmetrical configuration. The symmetrical divider was effective for operation in He, while the asymmetrical divider was more effective in Ar, Kr, and Xe.

In He at 1 atm, typical electric fields were $E_D \approx 0.6$ kV/cm in the drift gap, $E_T \approx 1.1$ kV/cm in the transfer gaps, $E_I \approx 2.2$ kV/cm in the induction gap; the voltage across a single GEM ("GEM voltage") was $\Delta V_{GEM} = 150\text{-}200$ V. In Kr at 1 atm, the corresponding values were $E_D \approx 1.0$ kV/cm, $E_T \approx 1.9$ kV/cm, $E_I \approx 1.9$ kV/cm, $\Delta V_{GEM} = 250\text{-}350$ V.

The anode signal was recorded either in a current or pulse-counting mode. The anode current value was always kept below 100 nA, reducing the X-ray tube intensity, to prevent charging-up of GEM foils. The maximum attainable gain was defined as that at which no anode current instabilities (dark currents or discharges) were observed for about 1 min.

In the current mode, the gain value of the triple GEM detector was defined as the anode current divided by the current induced by primary ionization in the drift gap. The latter current was determined in special measurements, where the drift gap was operated in an ionization mode.

In the pulse-counting mode, the gain value was determined with the help of a calibrated charge-sensitive amplifier: the anode charge was divided by the primary ionization charge produced in the drift gap. The latter charge was calculated using the data on X-ray absorption in an appropriate gas.

### 3. Detector performance in Kr

Fig.2 shows gain-voltage characteristics of the triple GEM detector in Kr, at different pressures, for symmetrical and asymmetrical divider configurations. One can see that the asymmetrical divider, with GEM voltage increasing towards the last GEM, allows to reach somewhat higher gains at high pressures.

It should be noted that in multi-GEM detectors using gas mixtures with molecular additives, the optimized GEM voltage decreased towards the last GEM [11], in contrast to operation in noble gases. This is probably because of the specific nature of the discharge mechanism in noble gases: the discharges are presumably generated by ion feedback from the last to preceding GEMs [7], due to an enhanced ion-induced electron emission as compared to other gases [12]. Decreasing the 1$^{st}$ GEM voltage would reduce this emission.

In general, the gain behaviour in Kr is very similar to that of Xe [10]: the maximum gain does not exceed $10^4$, weakly depending on pressure below 2 atm; at higher pressures it drops rapidly to below 10 at 5 atm. Moreover, similar to Xe, the operation in Kr turned out to be much more sensitive to discharges, compared to He, Ne and Ar. All three GEMs could be destroyed after even a few discharges when operating in Kr or Xe at maximum gains in the pressure range of 1-2 atm.

### 4. Detector performance in He

In Ne, Ar, Kr and Xe the primary ionization is generated by photoelectrons ejected from gas molecules due to X-ray photon absorption. In He, however, X-ray absorption in the gas is suppressed due to an extremely small absorption coefficient (see Appendix). This was confirmed by the following observation: in He the primary ionization current in the drift gap was practically independent from pressure, in contrast to other gases.

Apparently, the ionization in He is produced due to an X-ray-induced electron emission from solid, by the primary (energetic) and secondary (scattered) electrons. A typical depth of a solid from which a primary electron can be emitted without inelastic scattering is rather small, of the order of 20 atomic monolayers [13]. Nevertheless the calculations, presented in Appendix, show that this layer is thick enough to provide the signal. The electrons are emitted into the drift gap from the copper GEM electrode rather than from the aluminium cathode, since the photon absorption in Cu is by a factor of 30 larger than in Al.

The absorption of characteristic Mo K$_\alpha$ and Mo K$_\beta$ photons in the Cu K-shell would result in the ejection of primary electrons having the characteristic energies 8.5 keV and 10.7 keV, which are just the difference between the photon and K-shell energies. If the rest of the absorbed energy escapes detection, one would expect to see two peaks in the energy distribution. It should be remarked that the measurement of characteristic energies of electrons emitted from solid is the basic principle of ESCA (Electron Spectroscopy for Chemical Analysis) technique [13].

Fig.3 shows the anode pulse-height distribution in He at 5 atm at a gain of $9\times10^3$. Two peaks are distinctly seen. The relative peak positions correspond well to the electron characteristic energies. The more energetic part of the spectrum is presumably produced via Auger process, when the rest of the absorbed energy is released by ejection of additional electrons [13].

For each peak, one can estimate a total number of ion pairs created in the drift gap and thus calculate the detector gain, dividing the anode charge by this number. It is interesting that the energy resolution in He, estimated from the width of the peak, is close to that obtained with multi-GEM detectors in other, traditional gas mixtures [14]: $\sigma/E \approx 10\%$ at 8.5 keV.

Fig.4 shows anode signals after a charge-sensitive amplifier in He at 10 atm at a gain of $8\times10^3$. A strong line in the middle of the scale is distinctly seen, obviously corresponding to escape peaks considered above.

Gain-voltage characteristics in He at different pressures are shown in Figs. 5 and 6, in a current and pulse-counting mode, respectively. A symmetrical voltage divider, with an enhanced induction field, was used in these measurements. Comparing the results of two measurement techniques, one may conclude that they both give similar results and that the measurement uncertainty of the gain is within a factor of 2. The maximum gain in He weakly depends on pressure, reaching a value of $10^5$ at 15 atm.

Normally it is expected that the operation voltage of a gas detector substantially increases with pressure, as indeed was observed in heavy noble gases [10]. However, this is not the case for He and Ne. In He, in the pressure range of 1-7 atm, the operation voltage almost did not grow with pressure, increasing by only 10%. Moreover, in the range of 1-3 atm the operation voltage even decreased with pressure, at the initial part of the gain curve. To our knowledge, such unusual pressure dependence has never been observed before. This behaviour is very similar to that observed in Ne, where the operation voltage did not vary with pressure above 5 atm [10].

It should be noticed that the slopes of the gain curves are the same at all pressures, except in the final part of the curve for the data at 1 atm. The explanation is that at 1 atm the induction field was so high that it gave rise to a parallel-plate amplification mode in the induction gap, resulting in a stronger gain dependence on voltage. The onset of the parallel-plate mode at 1 atm was also indicated by a change in the anode signals: the pulse-height distribution became exponential. The parallel-plate mode was not observed at higher pressures.

We checked the data reproducibility, replacing the triple GEM detector with another one and adding a controlled amount of ambient air to He: of the order of $10^{-4}$ and $10^{-6}$. This is illustrated in Fig.7 showing the comparison of two sets of gain characteristics. The impurity test is of particular importance, since one should exclude from consideration the avalanche mechanism induced by impurities such as Penning effect. One can see that data are well reproduced.

5. Discussion

Fig.8 shows the dependence of the maximum gain of a triple GEM detector on pressure in all the gases studied. The difference between light (He, Ne) and heavy (Ar, Kr, Xe) noble gases is clearly seen. Together with the unusual behaviour of gain-voltage characteristics in He and Ne, this may indicate that a new avalanche mechanism arises at high pressures in light noble gases, other than the electron impact ionization.

We suggest a possible explanation to be related to the associative ionization mechanism [12,15-17]. In the associative ionization, the electron is produced in atomic collisions due to the association of an atom with an excited atom into a molecular ion: $He + He^* \rightarrow He^+_2 + e^-$. The energy threshold for this reaction is lower than that of the impact ionization, by about 1 eV [12,16,17]. The cross-section for the creation of molecular ions in He in this reaction is rather large, of the order of $10^{-15}$ cm$^2$ [12,15,17]. In addition, it is pointed out [16,18] that the impact ionization rate increases in proportion to the pressure ($p$), but the associative ionization rate with $p^2$. Therefore at high pressures the contribution of the associative ionization may exceed that of the impact ionization, at relatively low values of the reduced electric field. The detailed analysis of experimental results from this point of view is presented elsewhere [18]. It is interesting that according to theoretical calculations [19] the avalanche development in liquid He, at low electric fields, would also be defined by the associative ionization.

If we believe that the gas density is the main parameter defining the avalanche characteristics in noble gases, the highest gain of the multi-GEM detector would be achieved, at atmospheric pressure, at the following temperatures (derived from Fig.8): in Xe and Kr at 150K, in Ar at 100K, in Ne at 30K and in He below 20K. One can see that these temperatures are close to the boiling points of the appropriate gas. That means that the GEM structures could be successfully incorporated into double-phase cryogenic particle detectors.

## 6. Conclusions

We have studied the operation properties of a triple GEM detector in pure He, Kr and other noble gases at high pressures, varying from 1 to 15 atm. Light noble gases, He and Ne, provided a superior performance: the highest gain, approaching $10^5$, and remarkable gain dependence on pressure. The energy resolution of the triple GEM detector in compressed He was measured to be about 10% at 8.5 keV, which is close to that obtained in traditional gas mixtures.

In Ar, Kr and Xe the maximum gain rapidly drops for pressures exceeding 3 atm. In contrast, the maximum gain in He and Ne does not decrease with pressure. In addition, gain characteristics in He and Ne have an unusual pressure dependence: in a wide pressure range the operation voltage does not increase with pressure; moreover it can even start to decrease with pressure.

These results may indicate that a new avalanche mechanism starts playing a role at high pressures in light noble gases. We suppose that this mechanism is the associative ionization: at higher pressures it takes over the electron impact ionization due to stronger dependence on pressure and lower energy threshold. On the other hand, the associative ionization mechanism is not yet fully understood. In particular, it is not clear why it has a minor effect in heavy noble gases.

The results obtained are of high relevance for applications in cryogenic detectors for solar neutrino and dark matter search, where the operation of avalanche detector in noble gases at high gas densities is needed.

Other possible applications follow from the results obtained with the He-based detector. The high gain, good energy resolution and insensitivity to the direct ionization of the gas by X-rays of such a detector are very attractive for X-ray imaging and neutron detection with solid convertors [20]. The apparent application is a neutron detector using He$^3$ at high pressures, where He$^3$ would act as both a detecting and amplifying medium. Another possible application is the high-pressure helium Time Projection Chamber, proposed for solar neutrino detection [21]. Moreover, the adoption of noble gas as an amplifying medium, which does not age under avalanche conditions, offers a big advantage since it allows for operation in a sealed mode.

Further studies of this technique, e.g. GEM operation in pure noble gases at cryogenic temperatures, in a gas and liquid phase, at pressures higher than 20 atm are on the way.

## Acknowledgements

This work has been originally motivated by the possible application in cryogenic double-phase detectors for solar neutrino and dark matter search. We are indebted to Prof. W. Willis and Drs. J. Dodd and M. Leltchouk, of the Columbia University, and Dr. D. Tovey, of the Sheffield University, for having suggested these directions.

**Appendix**

Let us show that the signal due to the X-ray-induced electron emission in a He-based GEM detector is stronger than that of the X-ray absorption in He and that the emitted electrons are fully absorbed in the drift gap. The ionization process due to electron emission includes the following steps: absorption of 17.4 and 19.7 keV photons in the Cu electrode of the 1st GEM, emission of 8.5 and 10.7 keV primary (energetic) electrons from Cu into the drift gap and absorption of these electrons in the gap. The ionization efficiency per incident photon for this process can be estimated using the mean free path for inelastic scattering of the primary electron in Cu, $\lambda_e(Cu)$, and the X-ray absorption length in Cu, $\lambda_X(Cu)$: $\varepsilon(Cu) \approx \lambda_e(Cu)/\lambda_X(Cu)$. The value of $\lambda_e(Cu)$ for 10 keV electron is taken from [13]: it is equal to about 25 atomic monolayers or to $6.5 \times 10^{-7}$ cm, with an account of the Cu atomic diameter $a$=2.6 Å. For 20 keV photon, $\lambda_X(Cu)$=3.3×10$^{-3}$ cm [22]. Thus we have: $\varepsilon(Cu) \approx 2.0 \times 10^{-4}$.

The efficiency of ionization produced via X-ray absorption in He is: $\varepsilon(He) \approx d/\lambda_X(He)$, where $d$=0.3 cm is the drift gap thickness, $\lambda_X(He)$=2.9×10$^4$ cm is the absorption length for 20 keV photon in He at 1 atm [22]. Thus we have at 1 atm: $\varepsilon(He) \approx 1.0 \times 10^{-5}$ and $\varepsilon(Cu)/\varepsilon(He) \approx 20$. That means that the contribution of the X-ray-induced electron emission from the GEM electrode is always bigger that of the X-ray absorption in He, even at higher pressures.

The range of primary electrons in He can be calculated using the formula [23]:
$R(mg/cm^2) = 412\, E^n$,
$n = 1.265 - 0.0954 \ln E(MeV)$
It gives for the range of 10 keV electron in He at 1 atm: $R(He)$=0.9 cm. Therefore, at pressures higher than 3 atm the primary electrons emitted from the GEM electrode, with the corresponding characteristic energies, are fully absorbed in the drift gap. This is reflected in two characteristic peaks seen in the pulse-height distribution.

15. B. Sitar, G. I. Merson, V. A. Chechin, Yu. A. Budagov, Ionization measurements, Springer, Berlin, 1993.
16. P. Rice-Evans, Spark, streamer, proportional and drift chambers, Richelieu, London, 1974.
17. B. M. Smirnov, Excited atoms, Energoizdat, Moscow, 1982 (in Russian).
18. A. Buzulutskov, Physics of multi-GEM structures, presented at the 8th Int. Conf. on Instr. for Colliding Beam Physics, Novosibirsk, 2002, to be published in Nucl. Instr. and Meth. A.
19. A. A. Belevtsev, Proceedings of the Int. Conf. on Liquid Radiation Detectors, Tokyo, 1992, p. 59.
20. A. Breskin, Nucl. Phys. B (proc. suppl) 44 (1995) 351.
21. P. Gorodetzky et al., Nucl. Instr. and Meth. A 433 (1999) 554.
22. O. F. Nemets, Y. F. Gofman, Reference book on nuclear physics, Naukova Dumka, Kiev, 1975 (in Russian).
23. Physical quantities, Eds.: I. S. Grigorjeva, E. Z. Meilikhova, Energoizdat, Moscow, 1991 (in Russian).

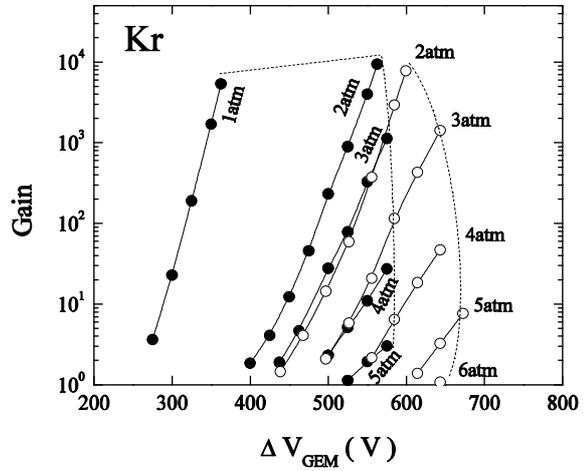

Fig.2 Gain of a triple GEM detector in Kr as a function of the voltage across the last GEM at different pressures, in a current mode. Two data sets are shown, for the symmetrical (solid points) and asymmetrical (open points) voltage divider.

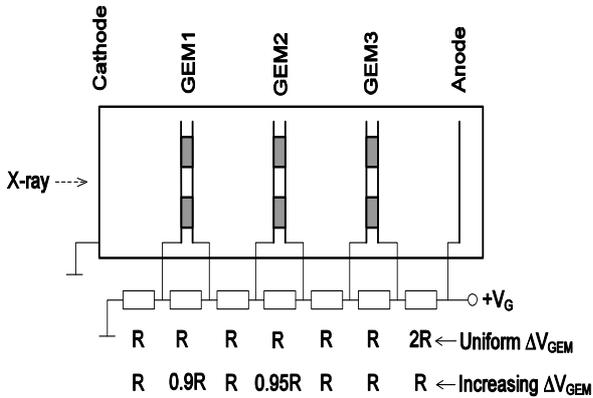

Fig.1 Schematic view of a triple GEM detector. Two voltage dividers were used, symmetrical and asymmetrical, with the uniform and increasing GEM voltage, correspondingly.

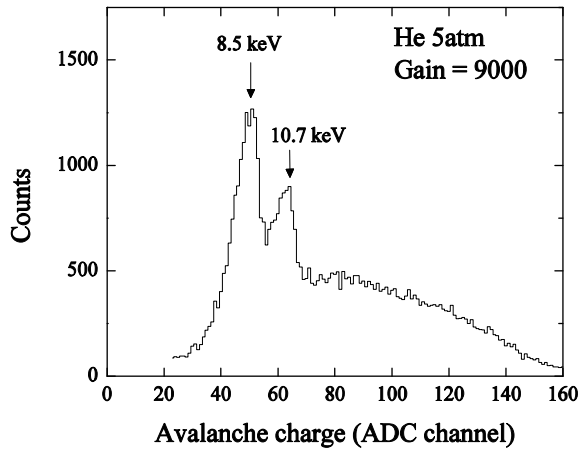

Fig.3 Pulse-height distribution of anode signals in He at 5 atm, at a gain of $9 \times 10^3$. Two peaks correspond to photoelectrons ejected from the Cu K-shell due to absorption of Mo $K_\alpha$ and Mo $K_\beta$ characteristic photons.

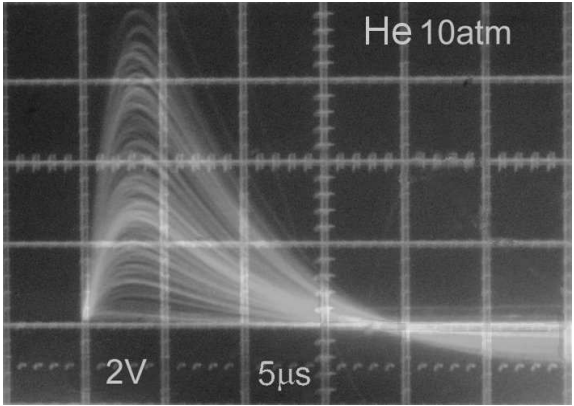

Fig.4 Anode signals in He at 10 atm, at a gain of $8\times10^3$, detected with a charge-sensitive amplifier.

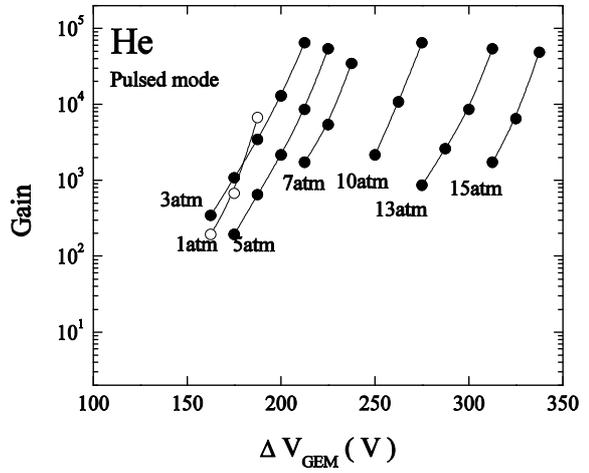

Fig.6 Gain of a triple GEM detector in He as a function of the voltage across each GEM at different pressures, in a pulse-counting mode.

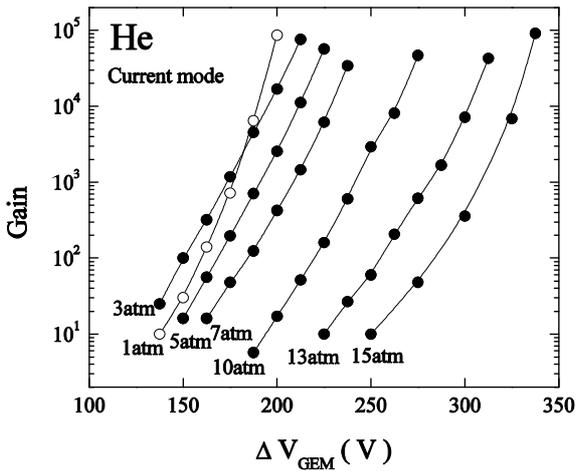

Fig.5 Gain of a triple GEM detector in He as a function of the voltage across each GEM at different pressures, in a current mode.

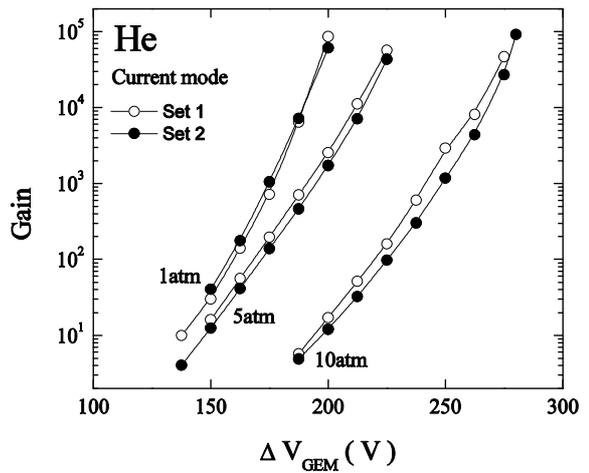

Fig.7 Data stability test in He. Two sets of gain-voltage characteristics are shown for two different triple GEM detectors and different air impurities, of the order of $10^{-4}$ (open points) and $10^{-6}$ (solid points).

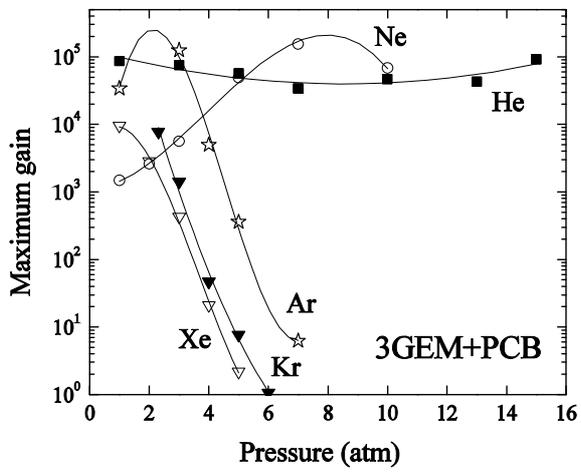

Fig.8 Maximum gain of a triple GEM detector as a function of pressure in He, Ne, Ar, Kr and Xe.